\documentclass[aps,reprint,pre,nofootinbib,showkeys]{revtex4-1}
\usepackage{amsfonts, amsmath, amssymb}
\usepackage[pdftex]{graphicx}
\usepackage{natbib}

\newcommand{\figFour}{Figure4.png}
\newcommand{\figFive}{Figure5.png}
\newcommand{\CC}{{\mathcal C}}

\newcommand{\mytitle}
    {Community-level cohesion without cooperation}
\newcommand{\myauthor}
    {Mikhail Tikhonov}
\newcommand{\myaffilA}
    {Center of Mathematical Sciences and Applications}
\newcommand{\myaffilB}
    {Harvard John A. Paulson School of Engineering and Applied Sciences}
\newcommand{\myaffilC}
    {Kavli Institute for Bionano Science and Technology, Harvard University, Cambridge, MA 02138, USA}
\newcommand{\myabstract}{Recent work draws attention to community-community encounters (``coalescence'') as likely an important factor shaping natural ecosystems. This work builds on MacArthur's classic model of competitive coexistence to investigate such community-level competition in a minimal theoretical setting. It is shown that the ability of a species to survive a coalescence event is best predicted by a community-level ``fitness'' of its native community rather than the intrinsic performance of the species itself. The model presented here allows formalizing a macroscopic perspective whereby a community harboring organisms at varying abundances becomes equivalent to a single organism expressing genes at different levels. While most natural communities do not satisfy the strict criteria of multicellularity developed by multi-level selection theory, the effective cohesion described here is a generic consequence of division of labor, requires no cooperative interactions, and can be expected to be widespread in microbial ecosystems.}

\newcommand{\vsigma}{{\vec\sigma}}

\newcommand{\EE}{{\mathcal{E}}}

\begin{document}
\title{\mytitle}
\author{\myauthor}
\email{tikhonov@fas.harvard.edu}
\affiliation{\myaffilA}
\affiliation{\myaffilB}
\affiliation{\myaffilC}
\begin{abstract}
\myabstract
\end{abstract}
\maketitle

Over the last decade, the sequencing-driven revolution in microbial ecology unveiled the staggering complexity of microbial communities that shape the health of our planet, and our own \cite{Caporaso11,Lozupone12,HMP,EMP}. These ecosystems routinely harbor hundreds of species of microorganisms, the vast majority of which remain poorly characterized. This makes the bottom-up approach to their modeling extremely challenging~\cite{Greenblum13,Bucci14,Boyang15}, prompting the question of whether some effective, top-down theory of the community as a whole might be a more viable alternative~\cite{Doolittle10,Borenstein12,Greenblum13,Bucci14}.

The need for a top-down approach is highlighted by multiple experimental observations. The microscopic species-level composition of independently assembled communities is highly variable even in similar environments; in contrast, the community metagenome (pathways carried by the population as a whole) appears to be more stable~\cite{HMP}. Studies of obesity or inflammatory bowel disease indicate that these conditions are unlikely to be caused by specific ``pathogenic species''~\cite{Major14,Mathur15}; similarly, the healthy human microbiota exhibits no core set of ``healthy'' microorganisms~\cite{HMP}. Thus, the ``healthy'' and ``diseased'' states of human-associated microbiota appear to be community-level phenotypic labels that may not always be traceable to specific community members.

Remarkably, the behavior of such macroscopically defined states can be productively studied even as the microscopic details remain unclear: thus, studies report on ``lean microbiota'' outcompeting ``obese microbiota'' in mice~\cite{Ridaura13}, or on the efficacy of fecal matter transplant in treating \emph{C. difficile} infections, whereby a ``healthy'' community overtakes the ``diseased'' state~\cite{Bakken11}. Both examples can be conceptualized as community-level competition events, termed ``community coalescence''. Although poorly understood, such events are widespread in natural microbial ecosystems and likely play a major role shaping their structure~\cite{Rillig15}. Intriguingly, Rillig~\emph{et~al.} argue that coalescing communities often appear to be ``interacting as internally integrated units rather than as a collection of species that suddenly interact with another collection of species''~\cite{Rillig15}.

Although comparing a community to a functionally integrated ``superorganism'' is a recurring metaphor~\cite{Shapiro98,West06}, a well-established body of theory cautions against using such terms loosely~\cite{Gardner09}. The formal criteria under which a group of organisms can be considered a ``multicellular whole'' have been extensively discussed in the context of multi-level selection theory (MLS)~\cite{Okasha08}. At the very least, the established notions of group-level individuality and ``organismality'' crucially rely on cooperative traits of group members~\cite{Buss87,Michod99,Michod03}. As a result, the formal applicability of the ``superorganism'' perspective appears to be severely restricted, as pervasive cooperation between members must first be demonstrated. In particular, the microbiota inhabiting the human gut is extremely unlikely to satisfy such criteria.

However, the utility of a macroscopic community-level perspective, and its ability to predict the outcome of competition between communities, need not hinge on whether they constitute a valid level of selection in the strict sense of MLS. It is well known that performance of a species is dependent on community context~\cite{Davis98,McGill06,McIntire14}: for example, niche-packed communities~\cite{MacArthur69,Roughgarden76} are more resistant to invasion\cite{Levine99}. Building on these ideas, the present work extends the classical model of MacArthur~\cite{MacArthur69} to construct a simple adaptive dynamics framework that describes co-evolution in multi-species communities~\cite{Roughgarden76,Geritz98,Nurmi08} and allows investigating the phenomenon of ``community coalescence'' in a minimal theoretical setting. The central result is a mathematically precise analogy established between a community whose members can change in abundance and an individual organism whose pathways can modulate in expression. This analogy concerns the manner in which a community interacts with its environment and with other communities; it does not investigate reproduction, and so does not constitute multicellularity in the established sense of the term~\cite{Okasha08}. Rather than being a limitation, this expands the potential applicability of the top-down perspective advocated here. While the criteria of ``true multicellularity'' are too stringent to apply to most natural communities, the phenomenon described in this work is a generic consequence of ecological interactions in a diverse ecosystem and requires no cooperative behavior or ``altruism''~\cite{Gardner09}.

\section{Methods: The metagenome partitioning model}
To investigate community coalescence in the simplest theoretical setting, consider the following model for division of labor in large communities. It is closely related to MacArthur's model of competitive coexistence on multiple resources~\cite{MacArthur69}; see Supplementary Material (SM).

\begin{figure}[t!]
\centering
 \includegraphics[width=0.95\linewidth]{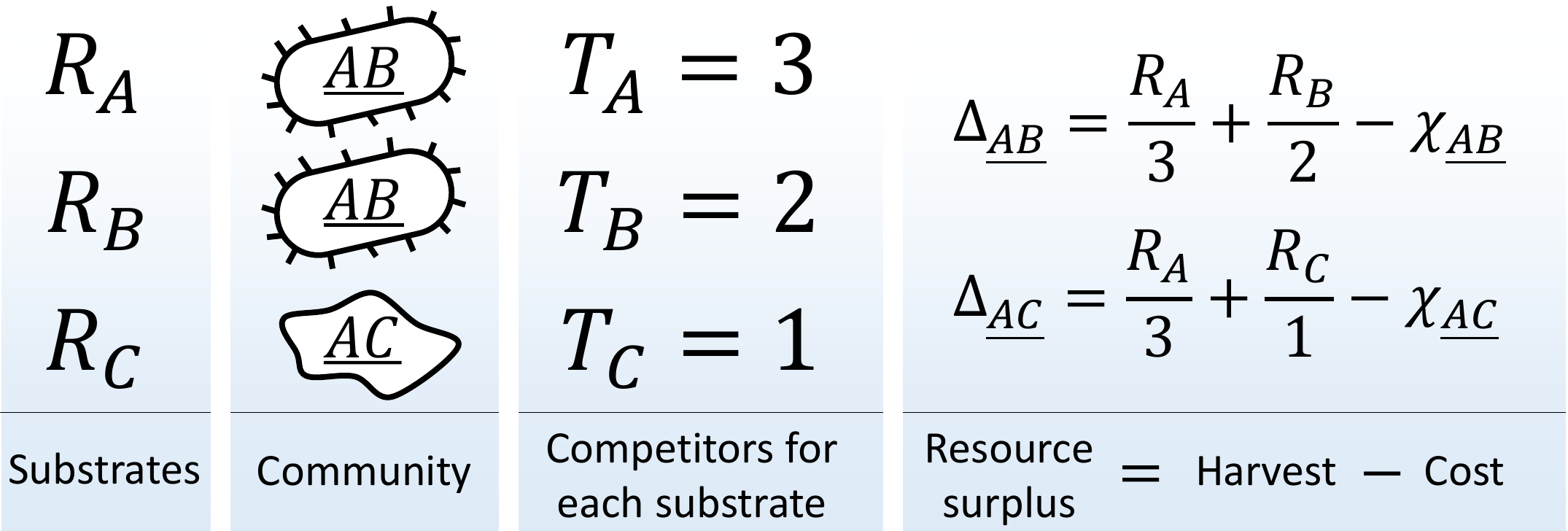}
\caption{\textbf{The metagenome partitioning model.} Organisms are defined by the pathways they carry, the benefit from each substrate is equally partitioned among all organisms who can use it, and population growth/death of each species is determined by the resource surplus it experiences.\label{fig:1}}
\end{figure}

Consider a community in a habitat where a single limiting resource exists in $N$ forms (``substrates'' $i\in\{1\dots N\}$) denoted $A$, $B$, etc. For example, this could be carbon-limited growth in an environment with $N$ carbon sources, or a community limited by availability of electron acceptors in an environment with $N$ oxidants. The substrates can be utilized with ``pathways'' $P_i$ (one specialized pathway per substrate). A species is defined by the pathways that it carries (similar, for example, to the approach of Ref.~\cite{Levin90}. There is a total of $2^N-1$ possible species in this model; they will be denoted using a binary vector of pathway presence/absence: $\vec \sigma=\{1,1,0,1,\dots\}$, or by a string listing all substrates they can use, e.g. ``species $\underline{ABD}$'' (the underline distinguishes specialist organisms such as $\underline{A}$ from the substrate they consume, in this case $A$). Let $n_\vsigma$ be the total abundance of species $\vsigma$ in the community, and let $T_i$ be the total number of individuals capable of utilizing substrate $i$ (Fig.~\ref{fig:1}):

$$
T_i\equiv \sum_{\vsigma} n_\vsigma \sigma_i.
$$
Assume a well-mixed environment, so that each of these $T_i$ individuals gets an equal share ${R_i}/{T_i}$ of the total benefit $R_i$ (carbon content, oxidation power; etc.) available from substrate $i$ (``scramble competition''). Any one substrate is capable of sustaining growth, but accessing multiple cumulates the benefits.
The population growth/death rate of species $\vsigma$ will be determined by the \emph{resource surplus} $\Delta$ experienced by each of its individuals:
\begin{equation}\label{eq:surplus}
\Delta_\vsigma = \sum_i \sigma_i \frac{R_i}{T_i} - \chi_\vsigma.
\end{equation}
Here the first term is the benefit harvested by all carried pathways, and the second represents the maintenance costs of organism $\vsigma$. These costs summarize all the biochemistry that makes different species more or less efficient at processing their resources. For simplicity, let these costs be random:
\begin{equation}\label{eq:cost}
  \chi_\vsigma = \chi_0|\vsigma|(1+\epsilon \,\xi_\vsigma).
\end{equation}
Here $\chi_0$ is a constant (the average cost per pathway), $\xi_\vsigma$ is a random variable chosen once for each species and drawn out of the standard normal distribution (truncated to ensure $\chi_\vsigma>0$), $\epsilon$ sets the magnitude of cost fluctuations, and $|\vsigma|\equiv\sum_i\sigma_i$ is the number of pathways carried by the species. This factor ensures that expressing more pathways incurs a higher cost (in this simple model, carrying and expressing a pathway is synonymous).

The resource surplus $\Delta$ is used to generate biomass. The simplest approach is to equate the biomass of an organism with its cost, so that the total biomass of a species is $\chi_\vsigma n_\vsigma$, and the dynamics of the model are given by:
\begin{equation}\label{eq:dynamics}
\tau_0\,\chi_\vsigma\frac{dn_\vsigma}{dt}\equiv g_\vsigma(\{n_\vsigma\})=n_\vsigma\Delta_\vsigma.
\end{equation}
The constants $\chi_0$ and $\tau_0$ set the units of resource and time.

%
%

The approach taken here purposefully ignores multiple factors, most notably trophic interactions or any other form of cross-organism dependence. This is intentional: it ensures that the interaction matrix

$$
M_{ab}\equiv \frac{\partial g_{\vsigma_a}}{\partial n_{\vsigma_b}}
$$
has no positive terms, i.e.\ the setting is purely competitive (indices $a$, $b$ label species). This helps underline that the whole-community behavior exhibited below is a generic consequence of division of labor, and requires no explicitly cooperative interactions.

Other simplifications include the assumption of deterministic dynamics and a well-mixed environment. Although stochasticity and spatial structure are tremendously important in most contexts, the simplified model adopted here provides a convenient starting point and makes the problem tractable analytically.

This work will investigate coalescence of communities that originate and remain in similar environments, e.g., transfer of oral communities by kissing~\cite{Kort14} as opposed to invasion of microbes from the mouth into the gut~\cite{Qin14}. Imagine a collection of islands (or patches) labeled by $\alpha$, each harboring a community $\CC_\alpha$ experiencing the same environment $\EE$. The next section investigates the within-island dynamics~\eqref{eq:dynamics} to establish some key properties that make this simplified model particularly convenient for our purposes. Specifically, let $\Omega(\CC)$ denote the set of species present at non-zero abundance in a community $\CC$. It will be shown that under the dynamics~\eqref{eq:dynamics}, any community $\CC$ will eventually converge to a stable equilibrium $\CC^*$ uniquely determined by the set $\Omega(\CC)$. Here and below, the starred quantities refer to equilibrium of ecological dynamics. At this equilibrium, certain species $S^*=\Omega(\CC^*)$ establish at a non-zero abundance, while others ``go extinct'', exponentially decreasing towards zero. Importantly, the set of survivors will depend only on the identity of the initially present species, and not on their initial abundance. Thus a community $\CC_1$ coalescing with $\CC_2$ will yield the same community $\CC_{12}^*$ irrespective of the initial mixing ratios. While obviously a simplification, this makes the metagenome partitioning model an especially convenient starting point to build theoretical intuition about community-community interactions before more general situations can be studied, e.g.\ numerically.

These properties are established in the next section; the following section then turns to the main subject of this work, namely coalescence events between islands.

\section{Single-island adaptive dynamics: intrinsic species performance and a community-level objective function}
%
%
\begin{figure}[b!]
\centering
 \includegraphics[width=0.95\linewidth]{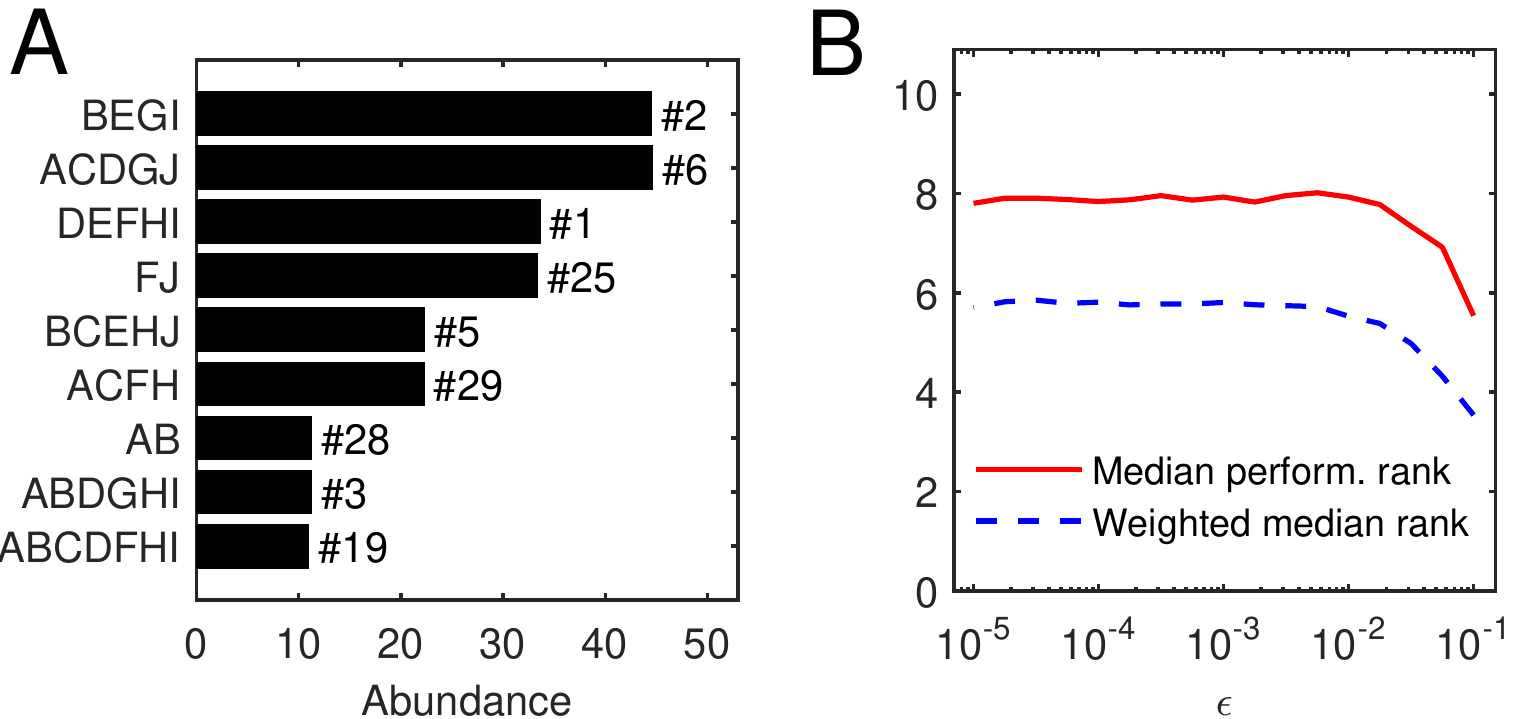}
\caption{\textbf{The individual performance rank of a species (its cost per pathway) is predictive of its survival and abundance in a community.} \textbf{A:}~Community equilibrium for one particular random realization of the model ($N=10$, $\epsilon=10^{-3}$). Species are ordered by abundance and labeled by the pathways they carry. Also indicated is the individual performance rank; all surviving species were within the top 30 (out of 1023). \textbf{B:} The median individual performance rank of survivors, weighted (dashed) or not weighted (solid) by abundance. Curves show mean over 100 random communities for each $\epsilon$; the standard deviation across 100 instances is stable at approximately 40\% of the mean for both curves, independently of $\epsilon$ (not shown to reduce clutter).\label{fig:2}}
\end{figure}

Numerical simulation of the competition between all 1023 possible species, initialized at equal abundance, for $N=10$, $\epsilon=10^{-3}$, $R_i=100\,\chi_0$, and one random realization of organism costs results in an equilibrium state depicted in Fig.~\ref{fig:2}A.\footnote{MATLAB scripts (MATLAB, Inc.) performing simulations and reproducing Figs.~2-5 are available upon request.} In this example it consists of 9 species. It is natural to ask: for a given initial set of competitors, what determines the species that survive?

In the present model, the only intrinsic performance characteristic of a species is its cost per pathway. Consider an assay whereby a single individual of species $\vsigma$ is placed in environment with no other organisms present, and, for simplicity, all substrates supplied in equal abundance $R_i=R$. The initial population growth rate in this chemostat is given by:

$$
\left.\frac{dn_\vsigma}{dt}\right|_{t=0}=\frac 1{\tau_0\chi_\vsigma}\left[\sum_iR_i\sigma_i-\chi_\vsigma\right]
=\frac 1{\tau_0}\left[R\frac{|\vsigma|}{\chi_\vsigma}-1\right]
$$
and abundance eventually equilibrates at $n_\vsigma=R|\vsigma|/\chi_\vsigma$. Both these quantities characterize performance of species $\vsigma$ (the term ``fitness'' is avoided as it is a micro-evolutionary concept that, strictly speaking, should be defined only within individuals of one species). Define the ``individual'' performance measure of species $\vsigma$ as
\begin{equation}\label{eq:f}
f_\vsigma\equiv \frac{|\vsigma|\chi_0}{\chi_\vsigma}-1.
\end{equation}
This definition is convenient as it makes $f_\vsigma$ a dimensionless quantity of order $\epsilon$. Under the cost model~\eqref{eq:cost}, the performance ranking of species is random, set by the random realization of the costs $\xi$.

Predictably, this performance ranking is correlated with the success of a species in a community, but not very well (Fig.~\ref{fig:2}). The equilibrium depicted in panel~A predominantly consists of top-ranked species, and the median performance rank of surviving species is consistently low across a range of $\epsilon$ (panel~B). This median rank becomes even lower if the median is weighted by a species' abundance at equilibrium, indicating that top-ranked species tend to be present at higher abundance~\cite{Davis98,Birch53}. Still, at the equilibrium shown in Fig.~\ref{fig:2}A, the species ranked 4th in intrinsic performance went extinct, but 6 others ranked as low as \#29 remained present.

\begin{figure}[t!]
\centering
 \includegraphics[width=0.7\linewidth]{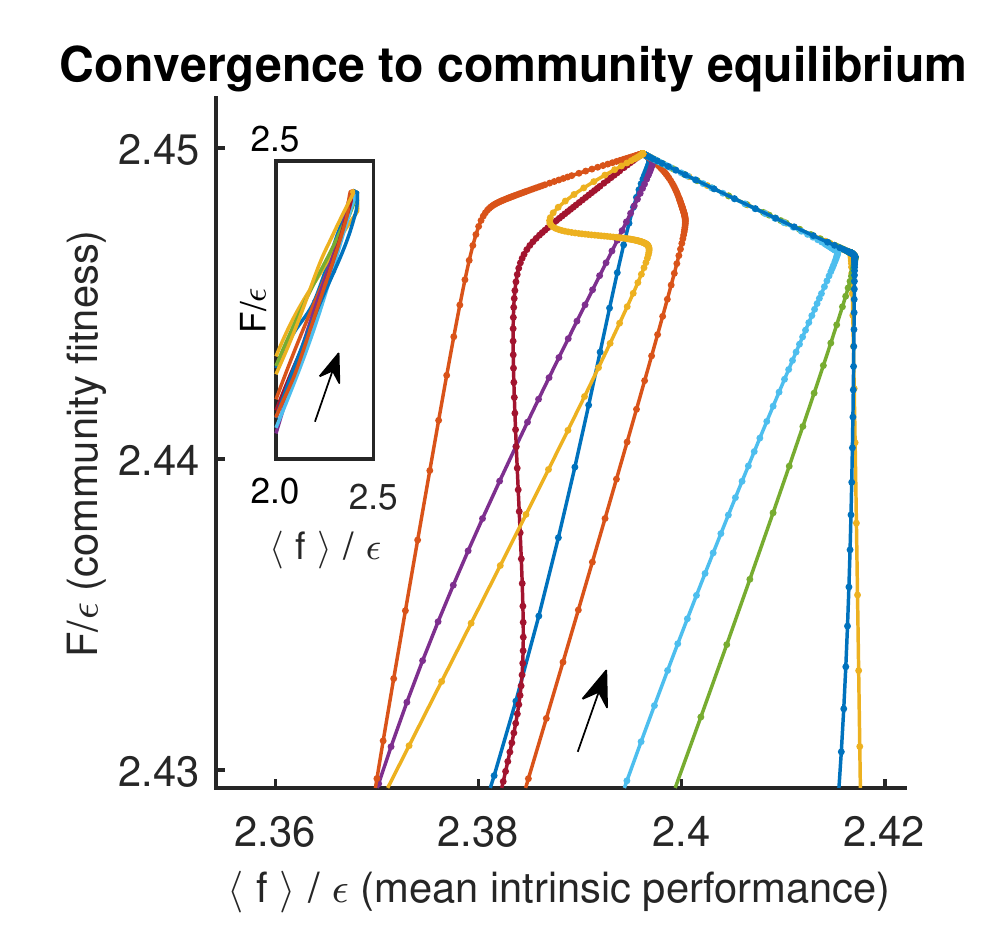}
\caption{\textbf{Community dynamics maximize a global objective function $F$.} 10 trajectories of ecological dynamics for an example system, starting from random initial conditions and converging to the equilibrium depicted in Fig.~\ref{fig:2}A. \textbf{Inset:} a zoomed-out version of the same plot; data aspect ratio as in the main panel. Mean intrinsic performance of community members is weighted by their abundance. Direction of dynamics indicated by arrows.\label{fig:3}}
\end{figure}

These observations reflect the well-known fact that the success of a species is context-dependent and observing a species in isolation does not measure its performance in the relevant environment~\cite{McGill06,McIntire14}. For example, consider the three-substrate world depicted in Fig.~\ref{fig:1}, and assume that $\underline{AB}$ is the highest-performing species with a very low cost. As $\underline{AB}$ multiplies, it depletes resources $A$ and $B$ (in the sense that the benefit $R_i/T_i$ any organism can harvest from them is reduced). As a result, the final equilibrium is highly likely to include the specialist organism $\underline{C}$, even if its cost is relatively high, and under other circumstances (if $\underline{AB}$ were less fit) it would have yielded to $\underline{AC}$ or $\underline{BC}$.
Conveniently, in the model described here, these complex effects studied by niche construction theory can be summarized in a single community-level objective function. The context experienced by all species is fully encoded in the vector of ``harvests'' $H_i\equiv R_i/T_i$ available from each substrate, and the dynamics~\eqref{eq:dynamics} possess a Lyapunov function (compare to MacArthur 1969):
\begin{equation}\label{eq:F}
  F=\frac 1{R_\text{tot}}\left(\sum_i R_i \ln \frac{T_i}{R_i/\chi_0} - \sum_\vsigma\chi_\vsigma n_\vsigma+R_\text{tot}\right).
\end{equation}
Here $R_\text{tot}$ is a constant introduced for later convenience. Specifically, set $R_\text{tot}=\sum_i R_i$; this choice ensures that close to community equilibrium, $F$ is also of order $\epsilon$ (see SM). This function, defined for $n_\vsigma\ge0$ and $T_i>0$, has the property that $R_\text{tot}\frac{\partial F}{\partial n_\vsigma}=\Delta_\vsigma$, and therefore

$$
\frac{dF}{dt} = \sum_\vsigma\frac{\partial F}{\partial n_\vsigma}\frac{dn_\vsigma}{dt} = \sum_\vsigma\frac{n_\vsigma\left(\Delta_\vsigma\right)^2}{R_\text{tot}\chi_0\tau_0}>0
$$
Thus $F$ is monotonically increasing as the system is converging to equilibrium. To illustrate this, Fig.~\ref{fig:3} shows 10 trajectories of ecological dynamics for the same system as in Fig.~\ref{fig:2}A, starting from random initial conditions (with all species present; see SM). Far from equilibrium, while most high-cost species are being eliminated by competitors, the mean intrinsic performance of surviving organisms and $F$ increase together (Fig.~\ref{fig:3}, inset), confirming that intrinsic performance is a useful predictor. However, as equilibrium is approached, community-induced changes in substrate availability $H_i$ reduce the relevance of the original performance ranking, which was measured in the ``wrong'' environment. The performance rank ordering will be all the more sensitive to the environment $H_i$, the smaller the scatter of intrinsic organism costs $\epsilon$. Therefore, the role of this parameter is to tune the relative magnitude of intrinsic and environment-dependent factors in determining a species' fate. So far, $\epsilon$ was fixed at $10^{-3}\approx 2^{-N}$, and Fig.~\ref{fig:2}B shows that for small $\epsilon$, the structure of the final equilibria does not significantly depend on this parameter (see SM). The large-$\epsilon$ regime will be discussed later.

%
%

Each of the trajectories in Fig.~\ref{fig:3} converges to the same equilibrium (depicted in Fig.~\ref{fig:2}A). This is because $F$ is convex and bounded from above (see SM). Therefore, for every set of species $\Omega$, any community restricted to these species will always reach the same (stable) equilibrium, corresponding to the unique maximum of $F$ on the subspace $V_\Omega$ defined by the conditions $\{n_\vsigma=0$ for all $\vsigma\notin\Omega\}$. This maximum will often be at the border of this subspace, corresponding to the extinction of some species.

Under the dynamics~\eqref{eq:dynamics}, no new species can ``appear'' if their original abundance was zero. Imagine, however, that on each island, a rare mutation (or migration) occasionally introduces a random new species; if it can invade, the community transitions to a new equilibrium and awaits a new mutation. This process of adaptive dynamics defines the evolution of each island, and can be seen as a mesoscopic population genetics model for a multi-species community evolving through horizontal gene transfer (loss/acquisition of whole pathways). For each island, $F$ is monotonically increasing throughout its evolution. Indeed, $F$ is continuous and non-singular in all $n_\vsigma$, so introducing an invader at a vanishingly small abundance will leave $F$ unchanged, and the subsequent convergence to a new equilibrium is a valid trajectory of ecological dynamics on which $F$ increases. Importantly, at any equilibrium,

$$
\sum_\vsigma\chi_\vsigma n_\vsigma=\sum_i R_i
$$
(see SM), and so the value of $F$ at community equilibrium is a quantity that depends \emph{only} on macroscopic quantities, namely the community-wide pathway expression $T_i$. The following sections will argue that $F$ can be thought of as community-level ``fitness'', but this term will not be used until justification is provided.

\section{The community-level function $F$ predicts the outcome of community coalescence}
%
%
\begin{figure*}[t!]
\centering
 \includegraphics[width=\textwidth]{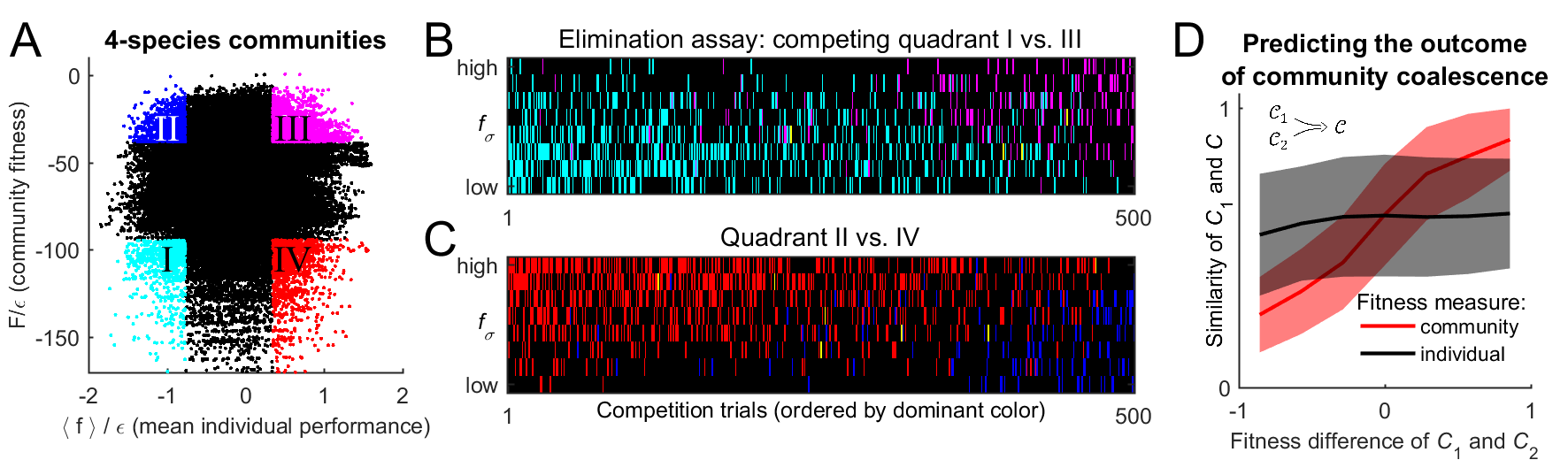}
\caption{\textbf{Community fitness is more predictive of competition outcome than the intrinsic performance of its members.} \textbf{A:}~Community fitness $F$ \emph{vs.}\ mean intrinsic performance $\langle f_\vsigma\rangle$ of its members, measured in units of $\epsilon$, for 70160 communities composed of 4 species (see text). Communities in which both characteristics are in the top or bottom 10\% are highlighted. \textbf{B:} Elimination assay competing quadrants I (cyan) vs III (magenta). 500 randomly drawn community pairs (columns) were jointly equilibrated, with up to 8 species each time (rows; ordered by $f_\vsigma$). For each species that went extinct during equilibration, the corresponding cell in the table is colored by the species' provenance. As expected, most eliminated species were from the less fit cyan communities (there are more cyan cells than magenta). These species also had lower $f_\vsigma$ (most colored cells are in the lower half of the table). \textbf{C:} Same, competing quadrants II (blue) vs IV (red). The dominant color is now red: most eliminated species were from red communities, and went extinct despite having higher $f_\vsigma$ (most colored cells are in the upper half of the table). Columns ordered by dominant color. \textbf{D:} Community similarity $S(\mathcal C_1, \mathcal C)$ for a coalescence event depicted in the cartoon (inset), computed for 5000 random community pairs, as a function of fitness difference between competing communities. Fitness difference scaled to the maximum of 1 so both fitness measures can be shown in same axes.  Shown is binned mean (8 bins) over communities with similar fitness difference (solid line) $\pm 1$ standard deviation (shaded).
\label{fig:4}}
\end{figure*}

Consider now a coalescence event whereby the equilibrium communities from two islands $\CC^*_\alpha$ and $\CC^*_\beta$ are brought into contact; as established above, the resulting community $\CC^*$ will not depend on the details of the mixing protocol. 
If none of the species from island $\beta$ could invade the community $\CC^*_\alpha$, then $\CC^*=\CC^*_\alpha$ and the community $\CC^*_\alpha$ is the clear winner. In general, however, the space of competition outcomes is richer than merely one community taking over: both competitors $\CC^*_\alpha$, $\CC^*_\beta$ can contribute to $\CC^*$, but can be more or less successful at doing so, contributing more or fewer species. What makes a community more likely to be successful?

The community on each island $\alpha$ constructs its own environment $\{H_i^{(\alpha)}\}$. When species from island $\alpha$ are introduced onto island $\beta$, they are exposed to a random new environment, and the equilibrium environment $\{H_i^*\}$ that the coalescence survivors will create for themselves will be different still. Although the success of a species is environment-dependent, for a random environment, $f_\vsigma$ as defined above remains the best available performance predictor. One may therefore expect that the more successful community should be the one with more high-performance species. On the other hand, we also found that the ultimate equilibrium community that cannot be invaded by \emph{any} species does not consist of species with the highest intrinsic performance, but corresponds to the global maximum of $F$. This suggests that the community-level function $F$ should be the better predictor of the competition outcome. If so, it could be said to characterize the ``collective fitness'' of a community (in the restricted, purely competitive, rather than reproductive, sense).

To settle the competition between these two hypotheses, the following procedure was implemented. For $N=10$, $\epsilon=10^{-3}$, and a given random realization of the cost structure $\xi$, $M=50$ random species were selected to allow for an exhaustive sampling of sub-communities (the results reported below do not significantly depend on this choice). This set was used to construct all $\binom{50}{4}=230300$ possible combinations of $k=4$ species that were independently equilibrated; instances where the equilibrium state had fewer than $k=4$ species or where some pathways were not represented were excluded. The putative collective fitness $F$ of the remaining 70160 communities, and the mean individual performance of their members, are shown in Fig.~\ref{fig:4}A. This procedure puts at our disposal multiple examples of communities where the two performance measures are both high, both low, or one is high while the other is low (the quadrants highlighted in Fig.~\ref{fig:4}A). Competing pairs of communities drawn from these pools will make it possible to determine which of the two factors, individual performance of a species $f_\vsigma$ or the collective fitness $F$ of its native community, can better predict its post-coalescence survival.

To begin, consider the competition between the cyan and magenta quadrants (I and III, respectively). Communities from the magenta quadrant are predicted to be more fit, both in the collective sense and as measured by the average intrinsic performance of members. Therefore, one expects that the magenta (III) communities should, on average, be more successful in pairwise competitions. To confirm this, Fig.~\ref{fig:4}B presents the results of an ``elimination assay'' competing communities from these quadrants. 500 random pairs were drawn, and correspond to columns in Fig.~\ref{fig:4}B. For each pair, species from both communities (up to 8 each time) were equilibrated together; the rows in Fig.~\ref{fig:4}B correspond to these species, ordered by individual performance rank: high (top) to low (bottom). For each species that went extinct during equilibration, its provenance was identified (``did it come from the magenta or the cyan community?''), and the corresponding rectangle in Fig.~\ref{fig:4}B was colored accordingly; in the rare cases when the eliminated species was originally present in both communities, it was colored yellow. The dominant color in Fig.~\ref{fig:4}B is cyan, confirming that the cyan communities are typically less successful at contributing their members to the final equilibrium. Note also that the colored entries are predominantly located in the bottom half of the table: the eliminated species tend to also have lower intrinsic performance than their more successful competitors. This is the expected result.

Now, consider the competition between blue and red quadrants (II and IV). An elimination assay conducted in an identical manner is presented in Fig.~\ref{fig:4}C. Now the colored entries are predominantly red and occupy the \emph{top} half of the table. In other words, members of the red communities are being outcompeted despite the fact that their intrinsic performance is higher: the individual performance of a species is less predictive of its ability to survive coalescence than the collective fitness of the community of which it was part.

Finally, 5000 random community pairs from the pool of Fig.~\ref{fig:4}A (not restricted to any quadrant) were competed. Define community similarity for $\mathcal C_1\equiv\{n_{1\vsigma}\}$ and $\mathcal C_2\equiv\{n_{2\vsigma}\}$ as the normalized scalar product of their species abundance vectors:

$$
S(\mathcal C_1, \mathcal C_2)=\frac{\sum_\vsigma n_{1\vsigma} n_{2\vsigma}}
{\sqrt{\sum_\vsigma n_{1\vsigma}^2}\sqrt{\sum_\vsigma n_{2\vsigma}^2}}.
$$
For each of the 5000 coalescence instances $\mathcal C^*_1+\mathcal C^*_2\mapsto \mathcal C^*$, Fig.~\ref{fig:4}D plots the similarity $S_{1}\equiv S(\mathcal C^*_1,\mathcal C^*)$ as a function of fitness difference between ``parent'' communities $\mathcal C^*_1$ and $\mathcal C^*_2$. It comes as no surprise (\emph{cf.} Fig.~\ref{fig:2}) that the predictive power of the mean individual performance is extremely weak (black line). In contrast, community fitness is a strong predictor:
the larger the difference in community fitness, the stronger the similarity between the post-coalescence community and its more fit parent (red line). In the mathematical framework developed here, the observation that coalescing communities appear to be ``interacting as coherent wholes'' acquires a precise formulation. Without implying the emergence of any new level of selection, and without invoking any cooperative traits, we observe that community coalescence can be usefully described as an interaction between two entities, characterized macroscopically at the whole-community level.

\section{The ``community as an individual'' metaphor becomes exact}
Consider now an external observer who is denied direct microscopic access to community composition, and is able to perform only ``metagenomic'' (or, rather, ``metaproteomic'') experiments, measuring the community-wide pathway expression $\vec T=\{T_i\}$ in response to substrate influx $\vec R=\{R_i\}$.

First, consider an island $\alpha_G$ harboring a single species: the complete generalist $\vsigma_G=\{1,1\dots 1\}$. Its abundance at equilibrium will be $n_G=T_i=R_{\text{tot}}/\chi_G$. Although substrates may be supplied in varying abundance, the island $\alpha_G$ can only express all pathways at the same level.

Another island $\alpha_S$ might harbor a community of perfect specialists: $\underline{A}=\{1,0,0\dots\}$, $\underline{B}=\{0,1,0\dots\}$, etc. Faced with an uneven supply of substrates, this island will adjust expression levels $T_i$ to precisely track the supply vector $R_i$, so that $T_i= R_i/\chi_i$, where $\chi_i$ is the cost of the respective specialist. For an external observer whose toolkit is limited to investigating the mapping $\vec R\mapsto \vec T$, the specialists' island $\alpha_S$ is formally indistinguishable from an organism who can sense its environment and up-regulate or down-regulate individual pathways.

Such perfect regulation is, however, costly: typically, $\underline{A}$, $\underline{B}$, etc.\ will not be the most cost-efficient combinations. As a result, allowing the community to evolve while holding $\vec R$ fixed, one will obtain a different multi-organism community $\CC$. Unlike $\alpha_S$, it will generally be unable to respond to all environmental perturbations: for example, the 9-species equilibrium community of Fig.~\ref{fig:2}A will necessariy be insensitive to some direction in the 10-dimensional space of substrate concentrations. Our external observer will conclude that evolution in a stable environment has traded some of the sensing capacity for the ability to fit a particular substrate influx with more efficient pathway combinations.

The model presented here can therefore be reinterpreted as a model for adaptive evolution of a single organism striving to better adjust its response $\vec T$ to the environment $\vec R$ it experiences. The model specifies how the genotype (patterns of pathway co-regulation) determines phenotype (the mapping $\vec R\mapsto \vec T$), and the competitive fitness $F$ as an explicit function of both the genotype and the environment~\cite{Ribeck15}. To conclude this section, let us compute the community fitness $F$ of the single-species generalist community $\alpha_G$ for the case $R_i\equiv R$. Applying the definition~\eqref{eq:F}, and using $T_i=n_G=NR/\chi_G$ one finds:

\begin{multline*}
F=\frac{1}{\sum_i R_i}\left(\sum_i R_i \ln \frac{T_i}{R_i/\chi_0} - n_G\chi_G\right) + 1 \\
 =\ln \frac{N\chi_0}{\chi_G} = \ln(1+f_G)\approx f_G
\end{multline*}
where $f_G$ is the individual performance~\eqref{eq:f} of organism $\sigma_G$, and the approximate equality holds because $f_G$ is of order $\epsilon$, assumed small. In other words, for a single-species community, the community fitness coincides with the individual performance of that species, reinforcing the emergent parallel between a community and an individual that had evolved an internal division of labor. This interpretation is specific to the particular model explored here, but within this model, the metaphor is mathematically exact.

\section{Community cohesion as a generic consequence of ecological interactions}
%
%
\begin{figure}[b!]
\centering
 \includegraphics[width=0.7\linewidth]{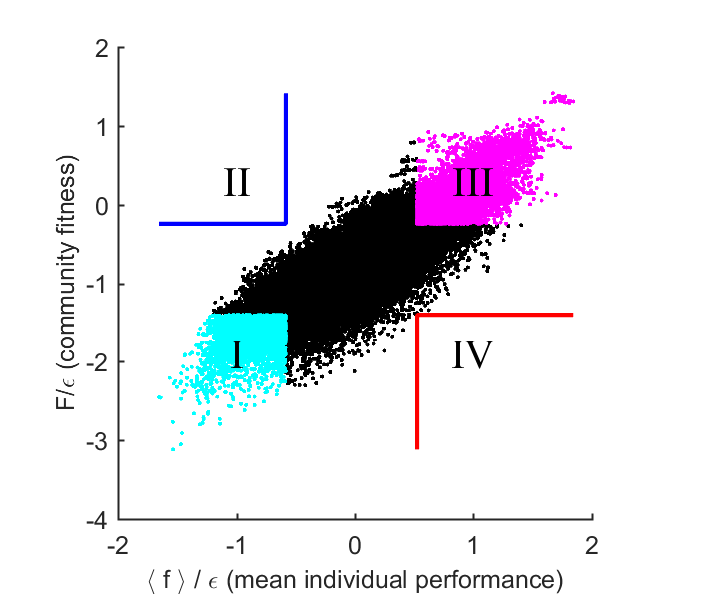}
\caption{\textbf{Parameter $\epsilon$ tunes the magnitude of community cohesion.} Same as Fig.~\ref{fig:4}A, for larger $\epsilon=0.1$. Increasing $\epsilon$ reduces the relative importance of environment in determining the performance ranking of species. As a result, collective fitness of a community and the mean individual performance of its members remains strongly coupled. Defining quadrants as in Fig.~\ref{fig:4}A leaves the blue and red quadrants empty.\label{fig:5}}
\end{figure}

It is important to contrast the results of the previous section with the notion of ``fitness decoupling'' in multi-level selection theory (MLS). In MLS, a higher level of organization is recognized when a group of cooperating organisms acquires interests that are distinct from the self-interest of its members~\cite{Okasha08}. Here, competition always remains entirely ``selfish''. In each instance of community competition assayed in Fig.~\ref{fig:4}, whenever some species invaded a community, it was because its fitness in \emph{that particular environment} was higher than the fitness of species already present. In contrast to fitness decoupling, which requires special circumstances to evolve, the community-level cohesion described in this work is a generic consequence of the fact that organisms modify their environment, and that fitness is context-dependent~\cite{Hay04,McGill06,McIntire14,Ribeck15}.

The definition~\eqref{eq:f} corresponds to how we might experimentally measure fitness, by placing an organism in a ``typical'' environment it is believed to experience. In the model described here, this typical environment is often an excellent approximation: for a community at equilibrium with equiabundant substrates $R_i = R$, the total community-wide expression of each pathway is roughly $T \approx R/\chi_0$, the same for all $i$. Nevertheless, even small deviations may be sufficient to induce substantial reordering of the relative performance rank of different species, in which case the context-dependent component of fitness can become dominant.

If this interpretation of the results of Fig.~\ref{fig:4} is correct, then reducing the degree to which environmental perturbations affect relative fitness of individuals should lead to a tighter link between community fitness and individual species' performance. This prediction can be tested by increasing $\epsilon$, the parameter that determines the width of the distribution of organism costs. For example, consider a community where the substrate $A$ is disputed by only two organisms: $\underline{A}$ and $\underline{AB}$. Assume that $f_{\underline{A}}>f_{\underline{AB}}$, so that when substrates $A$ and $B$ are equally abundant, the species $\underline{A}$ displaces $\underline{AB}$. Reducing the availability of substrate $A$ can reverse this outcome (if $A$ is absent, $\underline{AB}$ can still survive, but not $\underline{A}$). However, the larger the difference in intrinsic performance $f_{\underline{A}}$ and $f_{\underline{AB}}$, the more extreme such resource depletion would have to be. Therefore, increasing $\epsilon$ will reduce the relative effect that changing environment has on fitness rank ordering. Fig.~\ref{fig:5} repeats the analysis of Fig.~\ref{fig:4}A for $\epsilon=0.1$ (rather than $\epsilon=10^{-3}$ used previously). As predicted, the collective fitness is now strongly associated with the performance of individuals. In fact, this is already apparent in Fig.~\ref{fig:2}B: as $\epsilon$ is increased, the median fitness rank of survivors at the final equilibrium begins to reduce. At high $\epsilon$, it is increasingly true that high collective fitness is merely a reflection of high intrinsic performance of community members. Thus Fig.~\ref{fig:2}B documents a transition between a largely individualistic regime (at large $\epsilon$) and a regime where the genetically inhomogenous assembly of species increasingly acts ``as a whole'', in the precise sense discussed in the previous section.

\section{Discussion}
This work presented a theoretical framework where the analogy between a community harboring organisms at varying abundances, and an organism expressing genes at different levels, becomes an exact mathematical statement. A striking feature of this perspective is the blurred boundary between the notions of competition and genetic recombination ~\cite{Shapiro12,Rosen15}. Consider competition between organisms as an operation that takes two organisms and yields one:

\begin{equation*}
\text{Competition:}\quad (\mathcal O_1, \mathcal O_2)\mapsto \mathcal O_*.
\end{equation*}
Traditionally, the space of outcomes is binary: one competitor lives, one dies, and the propensity to survive competition is called fitness. When competition between communities of organisms is considered, this definition must inevitably be generalized to allow $\mathcal O_*$ to be distinct from either of the original competitors. Such ``competitors'', however, might be more aptly named ``parents''. In sexual reproduction, recombination allows a subset of the genes inherited from both parents to form progeny with potentially higher fitness; here, the competition between parent communities $\CC^*_\alpha$ and $\CC^*_\beta$ allows a subset of their members to regroup into a daughter community $\CC^*$ with a higher collective fitness $F$. The parallel becomes especially clear if one imagines propagules of $\CC^*_\alpha$ and $\CC^*_\beta$ co-colonizing a fresh environmental patch.

Such member regrouping can be much more flexible than the rules of sexual recombination, but reduces to the latter in the particular case of communities with clearly demarcated functional guilds (e.g.,\ consider competition between two communities that each has one plant, one pollinator, one herbivore, one carnivore, etc.). Long before the evolution of sex, such recombination would have allowed communities with divided labor to fix evolutionary novelty more efficiently than a clonal population of generalists. Although the metaphor of a genome as an ``ecosystem of genes'' is not new~\cite{Avise01}, the framework presented here allows it to be formalized and investigated quantitatively.

The results in this work were derived within the simplified framework of a particular model where microscopic dynamics conveniently took the form of optimizing a community-level objective function. In general, of course, collective dynamics are almost never reducible to solving an optimization problem~\cite{Metz08}. However, conceptually, the statement that environment-dependent species performance translates into an effective cohesion of coalescing communities is merely a generalization of the classical result that niche-packed communities are more resistant to invasion~\cite{MacArthur55}, which is recurrent across multiple modeling frameworks~\cite{Levine99}. In the model at hand, the existence of a global objective function made this phenomenon particularly easy to investigate; in a more general model, it wouldn't be possible to quantify this effect with a single number (the ``community fitness''). Nevertheless, the qualitative result may be expected to persist, so that members of a co-evolved community with a history of coalescence would tend to have higher persistence upon interaction with a ``na\"ive'' community that had never been exposed to such events, as proposed in Ref.~\cite{Rillig15}. More work is required to verify the generality of this hypothesis.

The results presented here, derived in a purely competitive model, demonstrate that functional cohesion is conceptually separate from the discussions of ``altruism'' and cooperation~\cite{Gardner09}, except to the extent described by the formula ``enemy of my enemy is my friend'' (indirect facilitation~\cite{Levine99b}). The latter can be seen as a form of cooperation~\cite{Hay04}, but is a generic phenomenon and is not vulnerable to ``cheaters''.

While the criteria of ``true multicellularity'' are too stringent to apply to most natural communities, the phenomenon described in this work is a generic consequence of ecological interactions in a diverse ecosystem. If whole-community coalescence events are indeed a significant factor shaping the evolutionary history of microbial consortia, then community-level cohesion of the type described here can be expected to be broadly relevant for natural ecosystems~\cite{Doolittle10}.

\section{Acknowledgments}
I thank Ariel Amir, Michael P.~Brenner, Andy Gardner, Jeff Gore, Miriam H.~Huntley, Simon A.~Levin, Anne Pringle, Ned S.~Wingreen and David Zwicker for helpful discussions, and anonymous referees for their comments on the early version of the manuscript. I have no competing interests. This work was supported by the Harvard Center of Mathematical Sciences and Applications, and the Simons Foundation.

\bibliographystyle{unsrtEtAl}
\bibliography{draft}

\onecolumngrid
 \appendix
 \cleardoublepage
 \section*{Supplementary material}
 \setcounter{figure}{0}
 \setcounter{equation}{0}
 \renewcommand{\theequation}{S\arabic{equation}}
 \renewcommand{\thefigure}{S\arabic{figure}}
 \renewcommand{\thetable}{S\arabic{table}}
\twocolumngrid

\subsection{Relation to the model of MacArthur}\label{sec:MacArthur}
The dynamics~\eqref{eq:dynamics} can be written as
\begin{equation}\label{eq:MacArthur}
  \frac{dn_{\vec\sigma}}{dt}=\frac1{\tau_0|\chi_\vsigma|} n_{\vec\sigma}\left(\sum_i \sigma_i A_i - \chi_\vsigma\right).
\end{equation}
where $A_i$ denotes the ``available resources''. In the model considered in this work, $A_i=\frac{R_i}{T_i}$. MacArthur (1969) considered a model of species competing for renewing resources. In that model, the dynamics of organism populations were identical to~\eqref{eq:MacArthur}, but the availability of resources was given by $A_i=R_i(1-T_i/r_i)$ (see equations (1)-(3) in MacArthur 1969), where the extra parameter $r_i$ is the renewal rate (or the ``intrinsic rate of natural increase'').

The dynamics of the two models, therefore, differ only by the choice of the functional form relating population growth and the corresponding decrease of resource availability. The mapping between the notations of MacArthur 1969 (``MA'') and those used here is provided in the table:
\begin{center}
\begin{tabular}{|l|c|c|}
  \hline
  Notation for... & MA & Here \\
  \hline
  Species index & $i$ & $\vsigma$\\
  Species abundance & $x_i$ & $n_\vsigma$ \\
  Resources a species can harvest & $a_{ij}$ & $\sigma_i$ \\
  Resource carrying capacity& $K_j$ & $R_i$ \\
  Minimal resource requirement& $T_i$ & $\chi_\vsigma$\\
  ``Resource weight'' & $w_i$ & 1\\
  Resources $\mapsto$ biomass conversion factor & $c_i$ & $(\tau_0\chi_\vsigma)^{-1}$\\
  Resource renewal rate & $r_j$ & N/A\\
  \hline
\end{tabular}
\end{center}

In the work of MacArthur, each species $i$ was described by an arbitrary chosen vector of parameters $a_{ij}$ (probability to encounter and consume resource $j$). The space of possibilities is unconstrained, and the types available to form a community are fixed by historical contingency; MacArthur then asks how many species can co-exist in this way. In the model considered here, $a_{ij}$ are constrained to be 0 or 1. The setting is treated as an adaptive dynamics model where species are allowed to acquire or lose pathways, and the outcome of this co-evolution is investigated.

Reformulating community dynamics as an optimization problem was first done in MacArthur 1969; here, because of the difference in the way resource consumption is treated, the objective function being optimized is different, but the argument is similar.
Consider the following objective function:
\begin{equation}\label{eq:Fnotnorm}
\tilde F=\sum_i R_i \ln T_i - \sum_\vsigma\chi_\vsigma n_\vsigma,
\end{equation}
defined for $\{n_\vsigma\ge0\}$, and differing from the definition of Eq.~\eqref{eq:F} only by normalization.

\emph{$\tilde F$ is bounded from above.}
To see this, note the inequalities:
$$
\sum_i T_i=\sum_\vsigma |\vsigma|n_\vsigma\le N\sum_\vsigma n_\vsigma
$$
and for $\alpha,\beta>0$:
$$
\alpha\ln x-\beta x\le \alpha\ln\frac{\alpha}{e\beta}
$$
Using these, and setting $\min_\vsigma\chi_\vsigma=\chi^*>0$, one can write:
\begin{multline*}
  \tilde F\le \sum_i R_i\ln T_i - \chi^* \sum_\vsigma n_\vsigma \le \sum_i \left(R_i\ln T_i - \frac {\chi^*}{N} T_i\right)\\
  \le\sum_i R_i\ln \frac{NR_i}{e\chi^*}
\end{multline*}

\emph{$\tilde F$ is convex.}
To see this, note that for any function $f(\vec n)$, the following two operations leave its convexity invariant ($M$ is an arbitrary matrix):
\begin{enumerate}
\item adding a linear function of its arguments:
    $$f(\vec n) \mapsto g(\vec n) = f(\vec n)+M\vec n;$$
\item performing a linear transformation of its arguments:
    $$f(\vec n) \mapsto h(\vec n) = f(M\vec n).$$
\end{enumerate}
Given these observations, convexity of $\tilde F$, and therefore also the convexity of $F$ as defined in~\eqref{eq:F}, directly follows from the convexity of the logarithm.

The main text demonstrated that $\tilde F$ is always increasing along the trajectories of the model. Thus, for any initial community state $\CC$, ecological dynamics converge to the equilibrium corresponding to the unique maximum of $\tilde F$ on the domain $\{n_\vsigma\ge0\text{ for }\vsigma\in\Omega(\CC)\}$. Since $\tilde F$ is bounded and convex, the final equilibrium always exists and is unique and stable.

\subsection{Normalization of community fitness}
The typical value of $\tilde F$ as defined in equation~\eqref{eq:Fnotnorm} for a community close to equilibrium can be estimated as follows.

To estimate the first term, note that the cost per pathway of all organisms is close to $\chi_0$, and therefore the overall expression $T_i$ is approximately $T_i\approx R_i/{\chi_0}$.

The second term is the total cost of all organisms in the population $\sum_\vsigma n_\vsigma\chi_\vsigma$. At any equilibrium, it is equal to the total resource abundance $R_\mathrm{tot}\equiv\sum_i R_i$. This can be seen in two ways. One approach is to use the equilibria conditions to express the cost of all present organisms in terms of resources:
$$
\forall \vsigma\in\Omega(\CC)\colon\chi_\vsigma=\sum_i \sigma_i \frac{R_i}{T_i}
$$
Therefore,
$$
\sum_\vsigma n_\vsigma\chi_\vsigma =
\sum_i \left(\sum_\vsigma n_\sigma \sigma_i\right) \frac{R_i}{T_i} = \sum_i R_i.
$$
Alternatively, this same equation can be derived from the condition of maximization of $\tilde F$, by setting $n_\vsigma\equiv M p_\vsigma$, and requiring $\frac{\partial \tilde F}{\partial M}=0$.

Putting these observations together, the expectation for the value of $\tilde F$ at any equilibrium is therefore
\begin{multline}
\tilde F=\sum_i R_i \ln T_i - \sum_\vsigma\chi_\vsigma n_\vsigma = \sum_i R_i \ln T_i - \sum_i R_i\\ \approx \sum_i R_i \ln (R_i/\chi_0) - \sum_i R_i \equiv \tilde F_0
\end{multline}
When defining community fitness, it is natural to subtract this baseline value from $\tilde F$ as defined in~\eqref{eq:Fnotnorm}, and normalize by $R_\mathrm{tot}$:
$$
F = \frac {\tilde F - \tilde F_0}{\sum_i R_i}.
$$
This is the normalization chosen in equation~\eqref{eq:F} in the main text.

\subsection{Sensitivity to the value of $\epsilon$}
Fig.~2B demonstrates that for small enough $\epsilon$, the structure of the final equilibria does not significantly depend on this parameter. This can be intuitively understood as follows. Consider two resources $A,B$ and organisms $\underline{A}=\{1,0\}$, $\underline{B}=\{0,1\}$, and $\underline{AB}=\{1,1\}$. If
\begin{equation}\label{eq:inequality}
  \chi_{\underline{AB}}>\chi_{\underline{A}}+\chi_{\underline{B}},
\end{equation}
it easily follows that the ``generalist'' organism $\underline{AB}$ will eventually be outcompeted by the two specialists $\underline{A}$ and $\underline{B}$. Conversely, if the opposite inequality holds, then $\underline{A}$ and $\underline{B}$ cannot stably coexist in the final equilibrium, since $\underline{AB}$ will always be able to invade, displacing one (or both) of them. In this way, in the metagenome partitioning model, community composition is shaped primarily by inequalities like~\eqref{eq:inequality}, which are invariant under changes in $\epsilon$ and depend only on the realization of the ``noise'' $\xi$.

\subsection{The maximum number of coexisting types}\label{sec:max}

The traditional question of how many types can coexist for a given set of parameters, although not at the focus of this work, is nevertheless instructive to address. A simple linear algebra argument demonstrates that in the model considered here, this maximum number is $N$: a stable coexistence is possible only for a number of types that is at most equal to the number of resources. This is because for a given set of $K$ types, the $K$ equilibria conditions $\Delta_\vsigma=0$ can be seen as a linear mapping between the $N$-dimensional vector $R_i/T_i$ and a $K$-dimensional vector of organism costs $\chi_\vsigma$. In the generic case (i.e. if no special symmetries exist in the cost structure), the existence of such a mapping requires $K\le N$.

Symmetries in the cost structure can lead to degenerate equilibria circumventing this maximal coexistence condition. Imagine, for example, that all organisms have the exact same cost per pathway $\chi_0$. In this maximally degenerate case \emph{any} combination of functional types can coexist, provided that $T_i=R_i/\chi_0$: no division of labor strategy is better than any other.

\subsection{Numerical determination of community equilibrium}
To determine the equilibrium state established through competition of a given set of $K$ species, one could imagine choosing a random starting point with a non-vanishing abundance of all $K$ competing species, and evolving it according to the dynamical equations for time $t\rightarrow\infty$. The Lyapunov function guarantees that such evolution would converge to an equilibrium state. However, if $K\gg N$ (for example, $K=1023$ and $N=10$ in Fig.~2A), such a procedure is highly memory-intensive and wasteful, since the final population is guaranteed to contain at most $N$ types with non-zero abundance (see section ``The maximum number of coexisting types'').

Conveniently, verifying that a configuration is a final equilibrium is much easier than finding it: one only needs to check that the resource surplus $\Delta_\vsigma$ is zero for all competitors that survived and is negative for all those who went extinct. This verification is fast and is guaranteed to either confirm that the equilibrium state is correct, or provide a list of species that can invade it. Therefore, a simple heuristic procedure can construct the true equilibrium configuration through an iterated sequence of ``guesses'', whereby a subset of species is first equilibrated, and then updated by removing species that went extinct and adding those that can invade. This is the approach adopted here.

Specifically, calculations were performed in Matlab (Mathworks, Inc.). Availability of all resources was set to $R=100$. The ``initial guess'' $S_0$ is constructed using the individual fitness criterion explained in the main text (low cost per pathway = high fitness): for each pathway $i$, the 10 most cost-efficient (lowest cost per pathway) functional types ($S_0^{(i)}$) that contained pathway $i$ are determined; the union of these cost-efficient types, all taken at equal abundance of 1 unit, constitutes the ``initial guess'' $S_0=\bigcup_i S_0^{(i)}$.

The following procedure is then iterated: community dynamics are simulated using MatLab's variable-order differential equation solver \verb"ode15s" until the absolute magnitude of all time derivatives $\frac{dn_\vsigma}{dt}$ fall below threshold $10^{-4}\epsilon$. At this point, most of the very-low-abundance species still present in the community are in the process of exponential extinction. To ensure that all such low-abundance types are indeed going extinct, all types with abundance below $10^{-4}$ are removed from the population, the pruned community is re-equilibrated (to account for any tiny adjustments this removal might have caused), and the resulting state $\CC^*$ is tested for being a non-invadable equilibrium. If any invaders are found, they are added to the community at abundance 1, and the simulation cycle is repeated. Otherwise (no species can invade), the configuration is accepted as being within the pre-determined numerical error of the true final equilibrium. This protocol ensures that in the community $\CC^*$, the list of survivors is exact (because the invadability criterion is always checked for all competing species and is exact), and their abundance is within acceptable numerical error. The protocol always converged due to convexity of ``community fitness'' $F$.

Scripts performing calculations and reproducing Figs.~2--5 are available upon request.

\subsection{Supplementary information for Figure 2B}

Figure~2B was generated as follows. For a given cost structure, 10 random subsets $\Omega_i$ of 100 types each were equilibrated to determine survivors $S^*_i$. The procedure was repeated for 10 random realizations of the cost structure at each $\epsilon$, with $\epsilon$ ranging from $10^{-5}$ to $0.1$. Thus for each value of $\epsilon$, a total of 100 randomly constructed communities were evaluated. Fig.~2B shows the median performance rank of survivors $S^*$ within the respective set of competitors, averaged over all 100 instances, where the median was either weighted (blue dashed line) or not weighted (red solid line) by abundance of the type at equilibrium.

\subsection{Initial conditions for Fig.~3}
The trajectories displayed in Fig.~3 were simulated for time $T=10^6$ starting from 10 random initial conditions whereby each of the 1023 types was set to an abundance value drawn out of a log-uniform distribution between $10^{-5}$ and $100$.

\end{document}